\documentclass[letterpaper,english,pra,aps,superscriptaddress,longbibliography,notitlepage,twocolumn]{revtex4-1}
\usepackage[T1]{fontenc}
\usepackage[latin9]{inputenc}
\usepackage{amsmath}
\usepackage{amssymb}
\usepackage{graphicx}

\makeatletter

\pdfpageheight\paperheight
\pdfpagewidth\paperwidth

\makeatother

\usepackage{babel}
\begin{document}
\preprint{}
\title{Anomalous Motion of Particle Levitated by Laguerre-Gaussian beam\\
}
\author{Yang Li}
\address{Quantum Physics and Quantum Information Division,~\\
 Beijing Computational Science Research Center, Beijing 100193, China}
\author{Lei-Ming Zhou}
\address{Quantum Physics and Quantum Information Division,~\\
 Beijing Computational Science Research Center, Beijing 100193, China}
\address{Department of Electrical and Computer Engineering, National University
of Singapore, Singapore 117583, Singapore}
\author{Nan Zhao}
\email{nzhao@csrc.ac.cn}

\address{Quantum Physics and Quantum Information Division,~\\
 Beijing Computational Science Research Center, Beijing 100193, China}
\begin{abstract}
Laguerre-Gaussian (LG) beam has orbital angular momentum (OAM). A
particle trapped in an LG beam will rotate about the beam axis, due
to the transfer of OAM. The rotation of the particle is usually in
the same direction as that of the beam OAM. However, we discovere
that when the LG beam is strongly focused, the rotation of the particle
and the beam OAM might be in the opposite direction. This anomalous
effect is caused by the negative torque on the particle exerted by
the focused LG beam, which is similar to the optical pulling force
in the linear case. We calculated the scattering force distribution
of a micro-particle trapped in an optical tweezers formed by the strongly
focused LG beam, and showed that there exist stable trajectories of
the particle that controlled by the negative torque. We proposed several
necessary conditions for observing the counter-intuitive trajectories.
Our work reveals that the strongly trapped micro-particle exhibits
diversity of motion patterns.
\end{abstract}
\maketitle

\section{\label{sec:level1}introduction}

Optical tweezers use radiation force of laser beams to trap or manipulate
micro- or nano-particles. Since the first discovery of laser trapping
of particles \citep{Ashkin1970,Ashkin1986,Chu1992}, optical tweezers
techniques have attracted more and more interest in the fundamental
research and applications, ranging from molecular biology to high-precision
measurement \citep{Marago2013,Gao2017}. Optical tweezers have become
an important tool for studying the non-equilibrium process of deoxyribonucleic
acid (DNA) in molecular biology \citep{Block1989,Kuo1992,Millen2014,Kasprzak2018}.
Recently, optical tweezers were also used to study the basic problems
of thermodynamics and to develop new types of quantum devices \citep{Li2012,Gieseler2014,Kim2019,Sompet2019}
.

The fundamental Gaussian mode of laser beams are the common choice
for the optical tweezers experiments \citep{Wright1993,Dienerowitz2008,Sinha2011}.
In recent years, laser beams with various non-trivial modes, such
as Bessel beams and Laguerre-Gaussian (LG) beams \citep{Arlt2001,Franke-Arnold2008,Marago2013,Monteiro2013,ZHANG-QI2013,Gong2016},
were introduced in the optical tweezers systems. Among those laser
beams, the LG beams, which carry orbit angular momentum (OAM) and
exert torque to the trapped particles \citep{Allen1992,Simpson1997,ZHANG-QI2013,Zhou2017},
can be used to manipulate the angular motion of the particles. Meanwhile,
with the angular motion considered, the particle exhibits remarkable
differences in trapping stability \citep{Ng2005,Ng2010}. Environmental
dissipation (friction) is essential to stabilize the angular motion
of the particle. 

In an optical tweezers system, the beams of non-trivial spatial modes
lead to many counter intuitive phenomena, such as the backward pulling
force generated by a forward propagating Bessel beam \citep{Chen2011,Ng2012,Chen2014,Novitsky2011,Kajorndejnukul2013}.
Moreover, the strongly focused laser beams exhibit different physical
characteristics from the paraxial beams. In the paraxial beam case,
an $\textrm{LG}_{pl}$ beam carries well-defined OAM with the quantum
number $l$. A particle trapped in the paraxial LG beam would feel
a torque in the same direction of OAM. However, when the LG beam is
strongly focused, the field distribution of the beam is significantly
modified. We found that the torque exerted on the particle is unnecessarily
the same as the OAM of the trapping beam. In the following, we refer
to the torque opposite to the OAM as the optical negative torque (ONT).

The ONT will significantly affect the angular motion of the trapped
particles. Driven by the ONT, the particle can rotate in the opposite
direction to the OAM of the trapping LG beam. A key question is whether
the anomalous trajectories is stable, so that they could be observed
experimentally. We used Lorentz-Mie theory to calculate force field
of the particle in the strongly focused laser beam and studied the
motion of the particle in the different dissipative environments.
We found that the dissipation caused by the residual pressure is crucial
to the trapping stability and the anomaly of motion. With suitable
dissipation, the anomalous trajectory of the particle driven by the
ONT would form a stable circle orbit. Besides, the orbits is robust
against to the perturbations on position and velocity of the particle.
We systematically study the dependence of the ONT on the laser intensity,
particle size and environmental dissipation, and propose a feasible
manipulation project in experiment.

\begin{figure}
\includegraphics[viewport=70bp 60bp 630bp 500bp,clip,width=7cm]{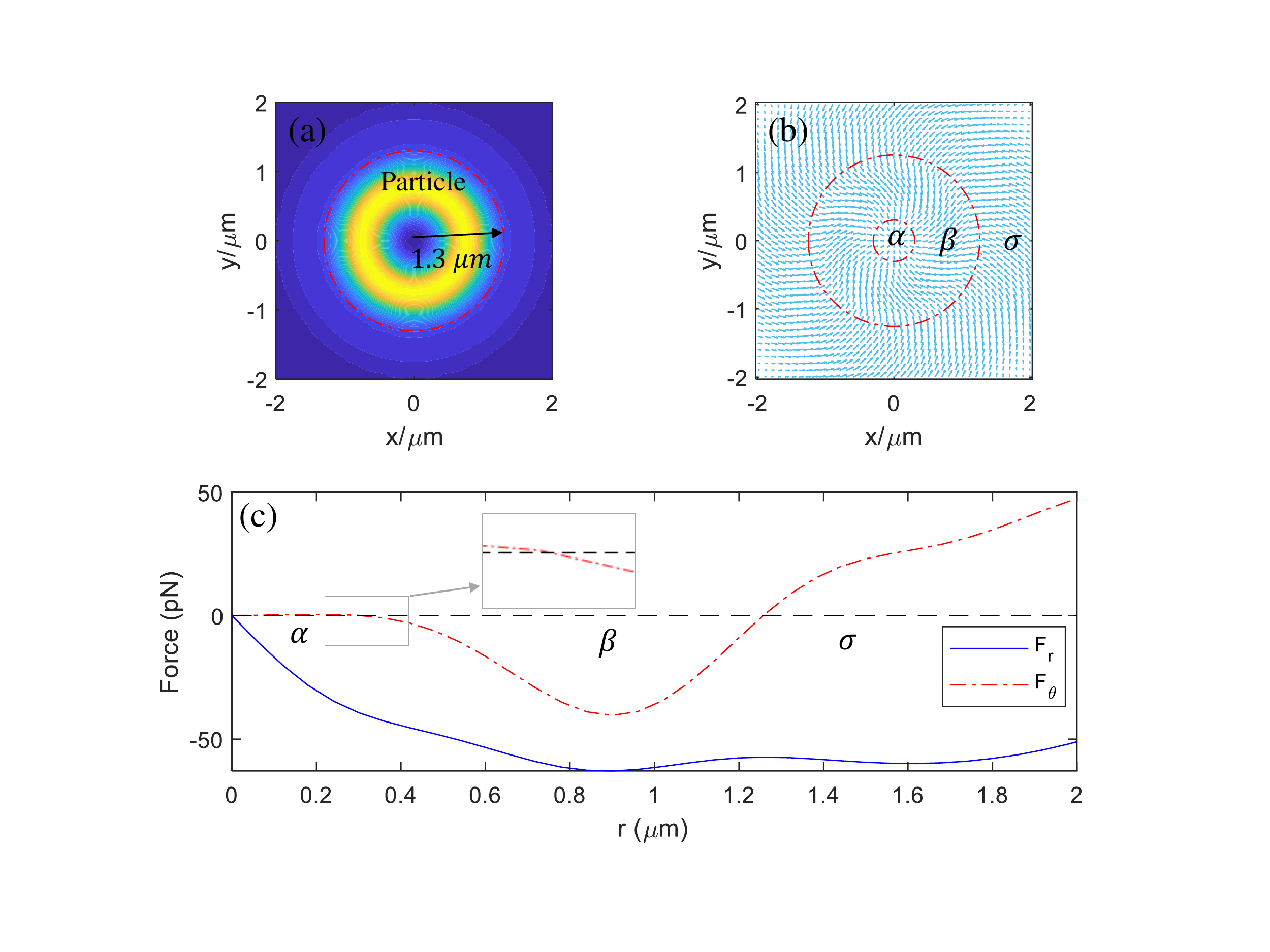}

\caption{(a) The intensity distribution of a strongly focused $\textrm{L\ensuremath{G_{03}}}$
beam in the focal plane. (b) The distribution of the radiation force
exerted on a particle of radius $R=1.3$ $\mu\textrm{m}$. (c) The
radial and azimuthal components of the radiation force, as functions
of the center-of-mass radius of the trapped particle. According to
the sign of the azimuthal force $F_{\theta}$, the focal plane can
be divided into three regions. In the second region $\beta$, the
direction of the optical torque is opposite to the direction of the
OAM of the beam.}
\end{figure}

\section{Force and Motion}

We consider a micro-particle which is trapped by strongly focused
Laguerre-Gaussian beams. Figure 1(a) shows the electric field distribution
of a typical circularly polarized $\textrm{LG}_{pl}$ beam with wavelength
$\lambda=1064$ $\textrm{\ensuremath{\mu}m}$ and the radical index
$p=0$ and azimuth index $l=3$. Because of the non-zero OAM, the
intensity distribution on the beam focal plane exhibits a bright ring.
Small particles with $R\ll\lambda$ are usually trapped on the bright
ring, which corresponding to a double-well \textquotedblleft optical
potential'' in the radial direction. However, for large particles
with $R\sim\lambda$ in our case, the center of the trapped particle
coincides the dark point of the beam \citep{Zhou2017}.

Figure. 1(b) shows the radiation force exerting on the center of mass
(CoM) of a glass spherical particle with refraction index $n=1.5$
and radius $R=1.3$ $\mu\textrm{m}$. The force field $F(r_{CoM})$
in the circularly polarized LG beam is rotational invariant. Figure
1(c) shows the radial component and azimuthal component, $F_{r}$
and $F_{\theta}$, respectively. The radial force is always negative
($F_{r}<0$) which indicates that the optical force tends to pull
the particle to the beam center. However, the azimuthal component
changes sign as increasing the radial distance $r_{cm}$ of the trapped
particle. There are three regions, according to the sign of the azimuthal
force {[}see Figs. 1(b)\&(c){]}. Particularly, the azimuthal force
become negative in the region with $0.3$ $\mu\textrm{m}$$<r<$$1.2$
$\mu\textrm{m}$. The optical torque $\tau$ exerted on the particle
is in the same direction of the OAM of the $\textrm{LG}_{03}$ beam
in the regions $\alpha$ and $\sigma$, while in the region $\beta$
the particle feels an ONT.

The counter-intuitive ONT causes non-trivial motion of the trapped
particle. We consider the equations of motion of the particle,
\begin{equation}
\begin{array}{c}
m\ddot{r}-mr\dot{\theta}^{2}=F_{r}-\gamma(P)\dot{r}\\
2m\dot{r}\dot{\theta}+mr\ddot{\theta}=F_{\theta}-\gamma(P)\dot{\theta}
\end{array},
\end{equation}
where m is the mass of the particle, $\gamma(P)$ is the friction
coefficient due to the residual gas. The transient motion relies on
the initial conditions of the particle (e.g., initial position and
velocity), which are usually not well-controlled. Here, we focus on
the stable rotation of the particle, which can be conveniently observed
in experiments. Due to the rotational symmetry of the radiation force,
an obvious solution of Eq. (1) is a uniform circular motion with $\dot{r}=0$
and $\dot{\theta}=0$. In this case, the radical force $F_{r}$ provides
the centripetal force, and the azimuthal force $F_{\theta}$ is compensated
by the friction force.

For a given pressure $P$, the two requirements in Eqs. (1)\&(1) determine
the possible radius $R^{*}$ of the circle orbit. In the low damping
limit, Eqs. (1)\&(1) have no solution, which means the friction is
too weak to slow down the angular motion of the particle. While in
the high damping limit, Eqs. (1)\&(1) only have trivial solution of
$R^{*}=0$. The particle, in this case, is confined around the beam
center. In the intermediate region, as shown in Fig. 2(a), there exists
several non-trivial solutions that the particle can have stable circular
orbits. More interestingly, some of the orbits lie in the ONT region,
where the angular velocity is opposite to the OAM of the trapping
beam. We refer to the counter-intuitive motion driven by the ONT of
the strongly focused LG beams as the anomalous motion of the particle
in the following.

To be observed in real experiments, the anomalous motion should be
robust against to perturbation. To analyze the robustness of the anomalous
motion, we expand Eq. (1) around the circle solution. With $r(t)=r_{0}+\delta r(t)$
and $\theta(t)=\theta_{0}+\delta\theta(t)$ and keeping up to the
linear terms of the fluctuations $\delta r(t)$ and $\delta\theta(t)$.
Using the normalization factor $\frac{1}{\gamma r_{0}}$, the Equation
(1) is linearized as

\begin{equation}
\left(\begin{array}{c}
\dot{a}\\
\dot{b}\\
\dot{c}
\end{array}\right)=\left(\begin{array}{ccc}
0 & \frac{\gamma}{m} & 0\\
\frac{m\omega^{2}+k_{1}}{\gamma} & -\frac{\gamma}{m} & 2\omega\\
\frac{k_{2}-\gamma\omega}{\gamma} & -2\omega & -\frac{\gamma}{m}
\end{array}\right)\left(\begin{array}{c}
a\\
b\\
c
\end{array}\right)
\end{equation}
with setting $a=\frac{1}{r_{0}}\delta\dot{r}$, $b=\frac{m}{\gamma r_{0}}\delta\ddot{r}$
and $c=\frac{m}{\gamma}\delta\ddot{\theta}$, where $\partial_{r}F_{r}=\partial F_{r}/\partial r|_{r=r_{0}}$
and $\partial_{r}F_{\theta}=\partial F_{\theta}/\partial r|_{r=r_{0}}$
are the derivatives of the radiation force components with respect
to the orbits radius. The eigenvalues $\{\lambda_{i}\}_{i=1}^{3}$
of the coefficient matrix determine the stability of the circle orbits.
A robust orbit requires that the real part of the eigenvalues should
be negative, i.e., $Re\mathrm{\mathsf{[\lambda_{i}]<0}}$ for $i=1$,
$2$ and $3$. Figure (2)b shows the maximal value of the real part
of three eigenvalues, i.e., $\mathcal{\varLambda}_{max}=max_{i}[Re[\lambda_{i}]]$,
as a function of the circular orbit radius. The particle can have
robust circle orbits with both positive and negative angular velocity. 

\begin{figure}
\includegraphics[viewport=80bp 30bp 650bp 500bp,clip,width=7cm]{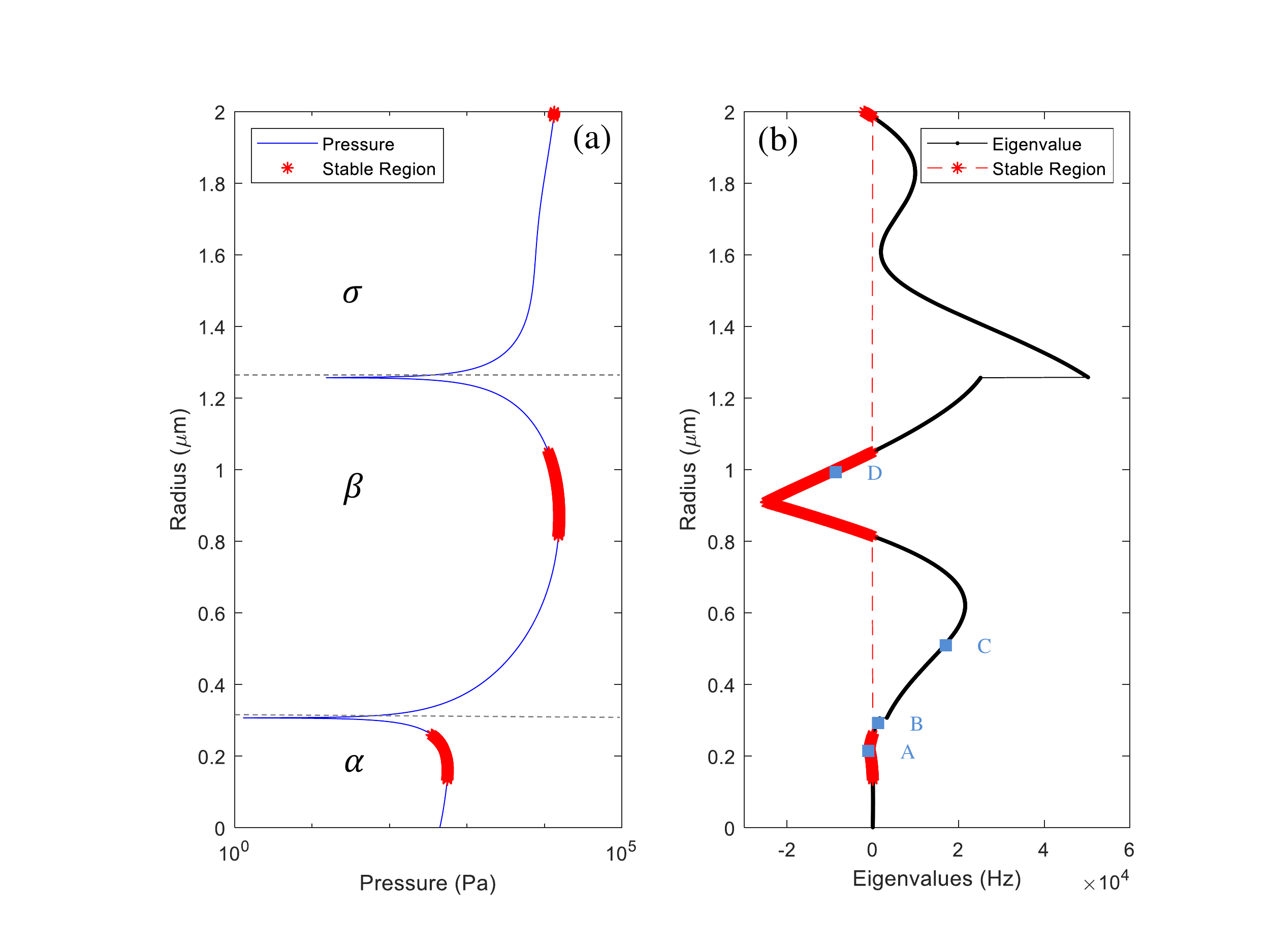}

\caption{(a) The circle orbits radius as a function of the given environmental
pressure $P$. Three regions are in the same as the regions defined
in Fig. (1). (b) The eigenvalue $\varLambda_{max}$ in the perturbation
analysis. The orbits with $\varLambda_{max}<0$ is stable (indicated
by red symbols). Four representative orbits are denoted by A - D,
which are clockwise stable orbit, clockwise unstable orbit, counter-clockwise
unstable orbit, and counter-clockwise stable orbit, respectively.}
\end{figure}

The dependence of radius of the robust circle orbits on the pressure
exhibits non-trivial behavior. In generic cases (e.g., trapping with
Gaussian mode beams), the radius of the circle orbit would be decreased
if one increases the environmental dissipation by increasing the pressure.
However, the inset of Fig. 2 shows an anomalous expansion of the orbit
size as increasing pressure. This is essentially caused by the non-trivial
angular force distribution of the strongly focused LG beam (see Fig.
1).

\begin{figure}
\includegraphics[viewport=120bp 50bp 620bp 500bp,clip,width=7cm]{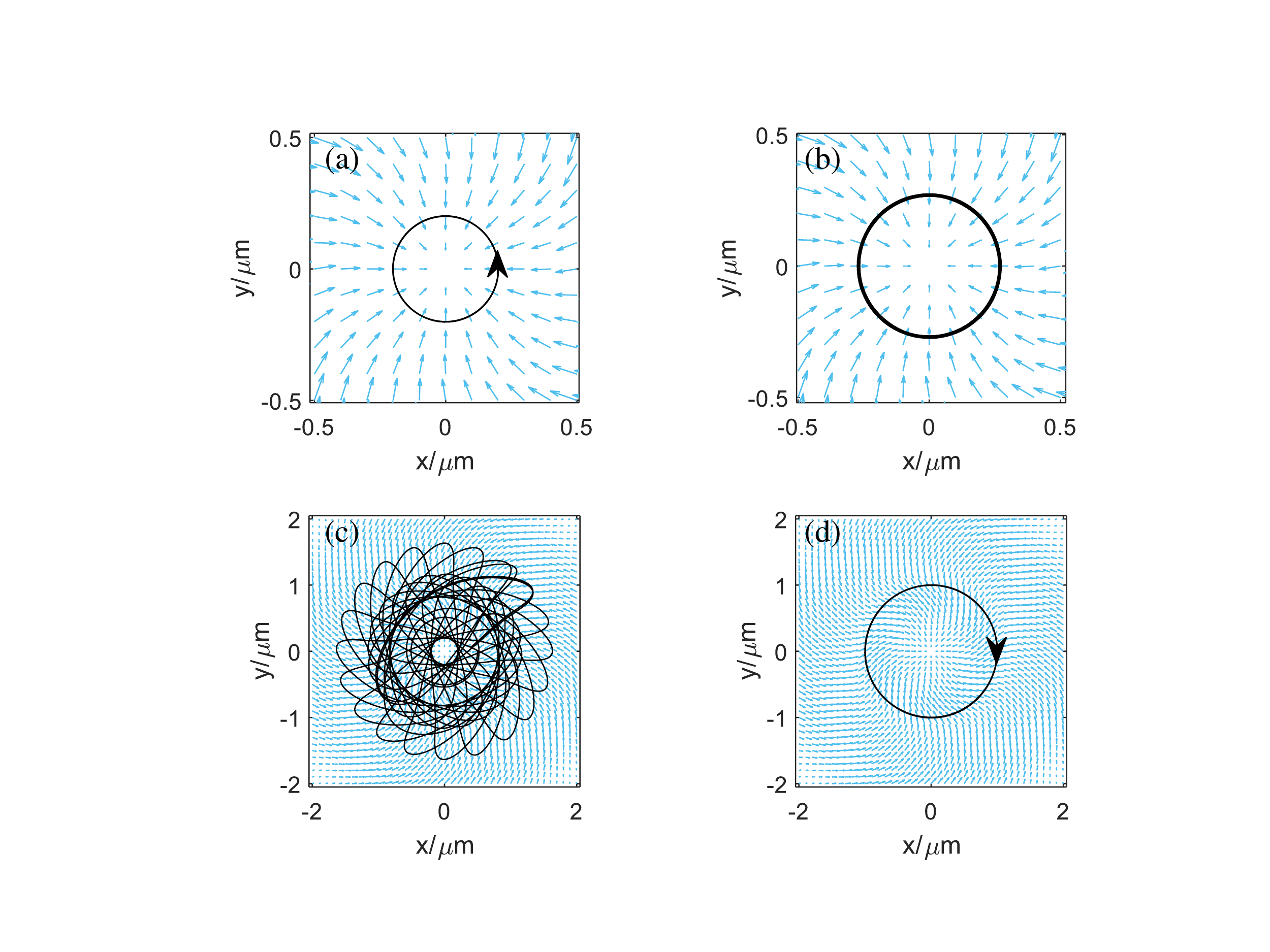}

\caption{The representative trajectories corresponding to the points A - D
in Fig. 2 under small perturbations. (a) The stable orbit corresponding
to point A. The angular velocity of the particle is the same as the
OAM of the beam. (b) \& (c) The unstable orbit corresponding to points
B and C. The radius of the particle oscillates in the trap. (d) The
stable orbit corresponding to point D. The angular velocity of the
particle is opposite to the OAM of the beam.}
\end{figure}

Figure 3 shows the typical trajectories of the trapped particle in
different regions under small perturbation. Given an environmental
pressure, e.g., $P=P_{D}$, the corresponding steady-state radius
and angular velocity of the circular motion is $r=r_{D}$ and $\dot{\theta}=\sqrt{F_{r}(r_{D})/mr_{D}}$,
where $P_{D}$ and $r_{D}$ are the coordinates of the point $D$
denoted in Fig. 2. Figure 3(a) shows the transient trajectory with
the initial radius perturbed by an amount of $1\%$, i.e., $\delta r/r_{A}=0.01$.
Since the point A is in the stable region, the initial perturbation
decays with time, and the particle is stabilized at the orbit with
radius $r=r_{A}$. The angular momentum is in the same direction of
the OAM of the trapping beam (counter-clockwise). However the same
amount perturbation around the unstable trajectories (e.g., the point
B and point C in Fig. 2) causes complex motions of the particle. For
small perturbation ($\delta r/r_{B}=\delta r/r_{C}=0.01$), the radius
$r(t)$ oscillates around the steady-state solutions $r_{B}$ and
$r_{C}$ {[}as shown in Figs. 3(b) and 3(c){]}, while strong perturbation
may cause the particle escape from the trap. Similar to Fig. 3(a),
Fig. 3(d) shows the stable trajectory in the anomalous region around
the point D in Fig. 2. Driven by the ONT, the particle has opposite
angular momentum direction (clockwise) to the OAM.

\begin{figure}
\includegraphics[viewport=20bp 0bp 650bp 480bp,clip,width=7cm]{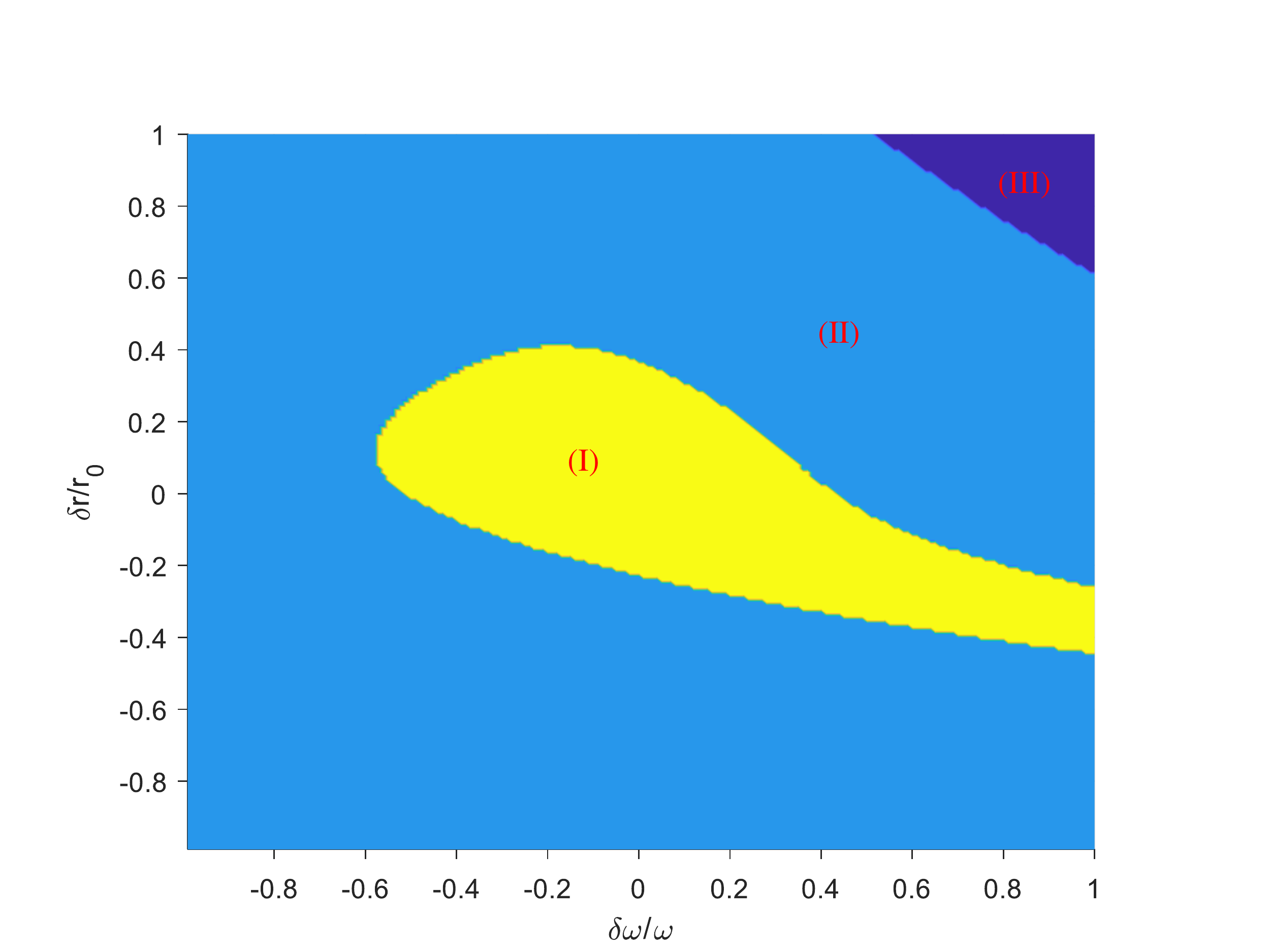}

\caption{Robustness of the anomalous motion under strong disturbance. The horizontal
axis is the relative change of the initial radius to the radius $r_{D}$
in Fig. 3(d). The vertical axis is the relative change of the initial
angular velocity to the angular velocity $\dot{\theta}_{D}$ in Fig.
3(d). In the yellow region (I), the particle is stabilized at the
anomalous orbit ($r=r_{D}$). In the blue region (II), the particle
is end up at the beam axis ($r=0$). In the purple region (III), the
disturbance is so strong that the particle escapes from the trap.}

\end{figure}

To further study the robustness of the anomalous motion in Fig. 3(d),
we calculate the particle's trajectory under strong disturbance. We
consider the initial radius and the initial angular velocity deviations
in a region of $\delta r/r_{D}=[-1,1]$ and $\delta\dot{\theta}/\dot{\theta}_{D}=[-1,1]$.
With different initial disturbance, the particle will end up with
three states in long term: (i) the stable anomalous orbit ($r=r_{D}$);
(ii) the beam axis ($r=0$); and (iii) escaping from the optical trap
($r=\infty$). Figure 4 shows that the particle can suffer quite strong
disturbance {[}as strong as $\sim40\%$ in both initial radius and
initial angular velocity, see the yellow region (I){]}. This result
indicates that the anomalous motion is robust, and is possible to
be observed even with not very precise control of the particle. 

\section{CONCLUSION}

In summary, we have studied the motion of a particle trapped in a
strongly focused LG beam. We predict that, similar to the optical
pulling force in the linear cases, the strongly focused beam can provide
counter-intuitive optical torques, i.e., the optical negative torques.
We further show that, the optical negative torques can drive the trapped
particle rotating in the opposite direction of the OAM of the trapping
LG beam. With the analysis of the motion stability, we conclude that
the anomalous motion driven by the ONT can be observed in experiments.

This work is supported by NKBRP (973 Program) 2016YFA0301201, the
Science Challenge Project (Grant No. TZ2018003), NSFC No. 11534002
and NSAF U1930402. L.-M. Z. acknowledges the support from Ministry
of Education, Singapore (Grant No. R-263-000-D11-114).

\bibliographystyle{plainnat}
\bibliography{REF}

\begin{thebibliography}{31}
\providecommand{\natexlab}[1]{#1}
\providecommand{\url}[1]{\texttt{#1}}
\expandafter\ifx\csname urlstyle\endcsname\relax
  \providecommand{\doi}[1]{doi: #1}\else
  \providecommand{\doi}{doi: \begingroup \urlstyle{rm}\Url}\fi

\bibitem[Allen et~al.(1992)Allen, Beijersbergen, Spreeuw, and
  Woerdman]{Allen1992}
L.~Allen, M.~W. Beijersbergen, R.~J.~C. Spreeuw, and J.~P. Woerdman.
\newblock Orbital angular momentum of light and the transformation of
  laguerre-gaussian laser modes.
\newblock \emph{Physical Review A}, 45\penalty0 (11):\penalty0 8185--8189, jun
  1992.
\newblock \doi{10.1103/physreva.45.8185}.

\bibitem[Arlt et~al.(2001)Arlt, Garces-Chavez, Sibbett, and Dholakia]{Arlt2001}
J~Arlt, V~Garces-Chavez, W~Sibbett, and K~Dholakia.
\newblock Optical micromanipulation using a bessel light beam.
\newblock \emph{Optics Communications}, 197\penalty0 (4-6):\penalty0 239--245,
  oct 2001.
\newblock \doi{10.1016/s0030-4018(01)01479-1}.

\bibitem[Ashkin(1970)]{Ashkin1970}
A.~Ashkin.
\newblock Acceleration and trapping of particles by radiation pressure.
\newblock \emph{Physical Review Letters}, 24\penalty0 (4):\penalty0 156--159,
  jan 1970.
\newblock \doi{10.1103/physrevlett.24.156}.

\bibitem[Ashkin et~al.(1986)Ashkin, Dziedzic, Bjorkholm, and Chu]{Ashkin1986}
A.~Ashkin, J.~M. Dziedzic, J.~E. Bjorkholm, and Steven Chu.
\newblock Observation of a single-beam gradient force optical trap for
  dielectric particles.
\newblock \emph{Optics Letters}, 11\penalty0 (5):\penalty0 288, may 1986.
\newblock \doi{10.1364/ol.11.000288}.

\bibitem[Block et~al.(1989)Block, Blair, and Berg]{Block1989}
Steven~M. Block, David~F. Blair, and Howard~C. Berg.
\newblock Compliance of bacterial flagella measured with optical tweezers.
\newblock \emph{Nature}, 338\penalty0 (6215):\penalty0 514--518, apr 1989.
\newblock \doi{10.1038/338514a0}.

\bibitem[Chen et~al.(2011)Chen, Ng, Lin, and Chan]{Chen2011}
Jun Chen, Jack Ng, Zhifang Lin, and C.~T. Chan.
\newblock Optical pulling force.
\newblock \emph{Nature Photonics}, 5\penalty0 (9):\penalty0 531--534, jul 2011.
\newblock \doi{10.1038/nphoton.2011.153}.

\bibitem[Chen et~al.(2014)Chen, Ng, Ding, Fung, Lin, and Chan]{Chen2014}
Jun Chen, Jack Ng, Kun Ding, Kin~Hung Fung, Zhifang Lin, and C.~T. Chan.
\newblock Negative optical torque.
\newblock \emph{Scientific Reports}, 4\penalty0 (1), sep 2014.
\newblock \doi{10.1038/srep06386}.

\bibitem[Chu(1992)]{Chu1992}
Steven Chu.
\newblock Laser trapping of neutral particles.
\newblock \emph{Scientific American}, 266\penalty0 (2):\penalty0 70--76, feb
  1992.
\newblock \doi{10.1038/scientificamerican0292-70}.

\bibitem[Dienerowitz(2008)]{Dienerowitz2008}
Maria Dienerowitz.
\newblock Optical manipulation of nanoparticles: a review.
\newblock \emph{Journal of Nanophotonics}, 2\penalty0 (1):\penalty0 021875, sep
  2008.
\newblock \doi{10.1117/1.2992045}.

\bibitem[Franke-Arnold et~al.(2008)Franke-Arnold, Allen, and
  Padgett]{Franke-Arnold2008}
S.~Franke-Arnold, L.~Allen, and M.~Padgett.
\newblock Advances in optical angular momentum.
\newblock \emph{Laser {\&} Photonics Review}, 2\penalty0 (4):\penalty0
  299--313, aug 2008.
\newblock \doi{10.1002/lpor.200810007}.

\bibitem[Gao et~al.(2017)Gao, Ding, Nieto-Vesperinas, Ding, Rahman, Zhang, Lim,
  and Qiu]{Gao2017}
Dongliang Gao, Weiqiang Ding, Manuel Nieto-Vesperinas, Xumin Ding, Mahdy
  Rahman, Tianhang Zhang, ChweeTeck Lim, and Cheng-Wei Qiu.
\newblock Optical manipulation from the microscale to the nanoscale:
  fundamentals, advances and prospects.
\newblock \emph{Light: Science {\&} Applications}, 6\penalty0 (9):\penalty0
  e17039, mar 2017.
\newblock \doi{10.1038/lsa.2017.39}.

\bibitem[Gieseler et~al.(2014)Gieseler, Quidant, Dellago, and
  Novotny]{Gieseler2014}
Jan Gieseler, Romain Quidant, Christoph Dellago, and Lukas Novotny.
\newblock Dynamic relaxation of a levitated nanoparticle from a non-equilibrium
  steady state.
\newblock \emph{Nature Nanotechnology}, 9\penalty0 (5):\penalty0 358--364, mar
  2014.
\newblock \doi{10.1038/nnano.2014.40}.

\bibitem[Gong et~al.(2016)Gong, Liu, Zhao, Ren, Qiu, Zhong, and Li]{Gong2016}
Lei Gong, Weiwei Liu, Qian Zhao, Yuxuan Ren, Xingze Qiu, Mincheng Zhong, and
  Yinmei Li.
\newblock Controllable light capsules employing modified bessel-gauss beams.
\newblock \emph{Scientific Reports}, 6\penalty0 (1), jul 2016.
\newblock \doi{10.1038/srep29001}.

\bibitem[Kajorndejnukul et~al.(2013)Kajorndejnukul, Ding, Sukhov, Qiu, and
  Dogariu]{Kajorndejnukul2013}
Veerachart Kajorndejnukul, Weiqiang Ding, Sergey Sukhov, Cheng-Wei Qiu, and
  Aristide Dogariu.
\newblock Linear momentum increase and negative optical forces at dielectric
  interface.
\newblock \emph{Nature Photonics}, 7\penalty0 (10):\penalty0 787--790, aug
  2013.
\newblock \doi{10.1038/nphoton.2013.192}.

\bibitem[Kasprzak et~al.(2018)Kasprzak, Kim, Le, Gao, Young, Yuan, Seog, Simon,
  and Shapiro]{Kasprzak2018}
Wojciech~K. Kasprzak, Taejin Kim, My-Tra Le, Feng Gao, Megan~Y.L. Young,
  Xuefeng Yuan, Joonil Seog, Anne~E. Simon, and Bruce~A. Shapiro.
\newblock Simulations of optical tweezers experiments reveal details of {RNA}
  structure unfolding.
\newblock \emph{Biophysical Journal}, 114\penalty0 (3):\penalty0 214a--215a,
  feb 2018.
\newblock \doi{10.1016/j.bpj.2017.11.1199}.

\bibitem[Kim et~al.(2019)Kim, Chang, Fields, Chen, and Hung]{Kim2019}
May~E. Kim, Tzu-Han Chang, Brian~M. Fields, Cheng-An Chen, and Chen-Lung Hung.
\newblock Trapping single atoms on a nanophotonic circuit with configurable
  tweezer lattices.
\newblock \emph{Nature Communications}, 10\penalty0 (1), apr 2019.
\newblock \doi{10.1038/s41467-019-09635-7}.

\bibitem[Kuo and Sheetz(1992)]{Kuo1992}
Scot~C. Kuo and Michael~P. Sheetz.
\newblock Optical tweezers in cell biology.
\newblock \emph{Trends in Cell Biology}, 2\penalty0 (4):\penalty0 116--118, apr
  1992.
\newblock \doi{10.1016/0962-8924(92)90016-g}.

\bibitem[Li et~al.(2012)Li, Gong, Yin, Quan, Yin, Zhang, Duan, and
  Zhang]{Li2012}
Tongcang Li, Zhe-Xuan Gong, Zhang-Qi Yin, H.~T. Quan, Xiaobo Yin, Peng Zhang,
  L.-M. Duan, and Xiang Zhang.
\newblock Space-time crystals of trapped ions.
\newblock \emph{Physical Review Letters}, 109\penalty0 (16), oct 2012.
\newblock \doi{10.1103/physrevlett.109.163001}.

\bibitem[Marag{\`{o}} et~al.(2013)Marag{\`{o}}, Jones, Gucciardi, Volpe, and
  Ferrari]{Marago2013}
Onofrio~M. Marag{\`{o}}, Philip~H. Jones, Pietro~G. Gucciardi, Giovanni Volpe,
  and Andrea~C. Ferrari.
\newblock Optical trapping and manipulation of nanostructures.
\newblock \emph{Nature Nanotechnology}, 8\penalty0 (11):\penalty0 807--819, nov
  2013.
\newblock \doi{10.1038/nnano.2013.208}.

\bibitem[Millen et~al.(2014)Millen, Deesuwan, Barker, and Anders]{Millen2014}
J.~Millen, T.~Deesuwan, P.~Barker, and J.~Anders.
\newblock Nanoscale temperature measurements using non-equilibrium brownian
  dynamics of a levitated nanosphere.
\newblock \emph{Nature Nanotechnology}, 9\penalty0 (6):\penalty0 425--429, may
  2014.
\newblock \doi{10.1038/nnano.2014.82}.

\bibitem[Monteiro et~al.(2013)Monteiro, Millen, Pender, Marquardt, Chang, and
  Barker]{Monteiro2013}
T~S Monteiro, J~Millen, G~A~T Pender, Florian Marquardt, D~Chang, and P~F
  Barker.
\newblock Dynamics of levitated nanospheres: towards the strong coupling
  regime.
\newblock \emph{New Journal of Physics}, 15\penalty0 (1):\penalty0 015001, jan
  2013.
\newblock \doi{10.1088/1367-2630/15/1/015001}.

\bibitem[Ng et~al.(2005)Ng, Lin, Chan, and Sheng]{Ng2005}
Jack Ng, Z.~F. Lin, C.~T. Chan, and Ping Sheng.
\newblock Photonic clusters formed by dielectric microspheres: Numerical
  simulations.
\newblock \emph{Physical Review B}, 72\penalty0 (8), aug 2005.
\newblock \doi{10.1103/physrevb.72.085130}.

\bibitem[Ng et~al.(2010)Ng, Lin, and Chan]{Ng2010}
Jack Ng, Zhifang Lin, and C.~T. Chan.
\newblock Theory of optical trapping by an optical vortex beam.
\newblock \emph{Physical Review Letters}, 104\penalty0 (10), mar 2010.
\newblock \doi{10.1103/physrevlett.104.103601}.

\bibitem[Ng et~al.(2012)Ng, Chen, Lin, and Chan]{Ng2012}
Jack Ng, Jun Chen, Zhifang Lin, and C.~T. Chan.
\newblock Pulling particles backward using a forward propagating beam.
\newblock \emph{The Journal of the Acoustical Society of America}, 131\penalty0
  (4):\penalty0 3533--3533, apr 2012.
\newblock \doi{10.1121/1.4709367}.

\bibitem[Novitsky et~al.(2011)Novitsky, Qiu, and Wang]{Novitsky2011}
Andrey Novitsky, Cheng-Wei Qiu, and Haifeng Wang.
\newblock Single gradientless light beam drags particles as tractor beams.
\newblock \emph{Physical Review Letters}, 107\penalty0 (20), nov 2011.
\newblock \doi{10.1103/physrevlett.107.203601}.

\bibitem[Simpson et~al.(1997)Simpson, Dholakia, Allen, and
  Padgett]{Simpson1997}
N.~B. Simpson, K.~Dholakia, L.~Allen, and M.~J. Padgett.
\newblock Mechanical equivalence of spin and orbital angular momentum of
  light:{\hspace{1em}}an optical spanner.
\newblock \emph{Optics Letters}, 22\penalty0 (1):\penalty0 52, jan 1997.
\newblock \doi{10.1364/ol.22.000052}.

\bibitem[Sinha et~al.(2011)Sinha, Köster, Ruez, Gonnord, Bastiani, Abankwa,
  Stan, Butler-Browne, Vedie, Johannes, Morone, Parton, Raposo, Sens, Lamaze,
  and Nassoy]{Sinha2011}
Bidisha Sinha, Darius Köster, Richard Ruez, Pauline Gonnord, Michele Bastiani,
  Daniel Abankwa, Radu~V. Stan, Gillian Butler-Browne, Benoit Vedie, Ludger
  Johannes, Nobuhiro Morone, Robert~G. Parton, Gra{\c{c}}a Raposo, Pierre Sens,
  Christophe Lamaze, and Pierre Nassoy.
\newblock Cells respond to mechanical stress by rapid disassembly of caveolae.
\newblock \emph{Cell}, 144\penalty0 (3):\penalty0 402--413, feb 2011.
\newblock \doi{10.1016/j.cell.2010.12.031}.

\bibitem[Sompet et~al.(2019)Sompet, Szigeti, Schwartz, Bradley, and
  Andersen]{Sompet2019}
Pimonpan Sompet, Stuart~S. Szigeti, Eyal Schwartz, Ashton~S. Bradley, and
  Mikkel~F. Andersen.
\newblock Thermally~robust spin correlations between two 85rb atoms in an
  optical microtrap.
\newblock \emph{Nature Communications}, 10\penalty0 (1), apr 2019.
\newblock \doi{10.1038/s41467-019-09420-6}.

\bibitem[Wright et~al.(1993)Wright, Sonek, and Berns]{Wright1993}
W.~H. Wright, G.~J. Sonek, and M.~W. Berns.
\newblock Radiation trapping forces on microspheres with optical tweezers.
\newblock \emph{Applied Physics Letters}, 63\penalty0 (6):\penalty0 715--717,
  aug 1993.
\newblock \doi{10.1063/1.109937}.

\bibitem[YIN et~al.(2013)YIN, GERACI, and LI]{ZHANG-QI2013}
ZHANG-QI YIN, ANDREW~A. GERACI, and TONGCANG LI.
\newblock {OPTOMECHANICS} {OF} {LEVITATED} {DIELECTRIC} {PARTICLES}.
\newblock \emph{International Journal of Modern Physics B}, 27\penalty0
  (26):\penalty0 1330018, oct 2013.
\newblock \doi{10.1142/s0217979213300181}.

\bibitem[Zhou et~al.(2017)Zhou, Xiao, Chen, and Zhao]{Zhou2017}
Lei-Ming Zhou, Ke-Wen Xiao, Jun Chen, and Nan Zhao.
\newblock Optical levitation of nanodiamonds by doughnut beams in vacuum.
\newblock \emph{Laser {\&} Photonics Reviews}, 11\penalty0 (2):\penalty0
  1600284, mar 2017.
\newblock \doi{10.1002/lpor.201600284}.

\end{thebibliography}

\end{document}